\documentclass[epj]{svjour}

\usepackage{graphicx}
\usepackage[utf8]{inputenc}
\usepackage[T1]{fontenc}
\usepackage[english]{babel}
\usepackage{braket}
\usepackage{multirow}
\usepackage{dsfont}
\usepackage{mathtools}
\usepackage{xcolor}
\usepackage[normalem]{ulem}

\newcommand{\beq}{\begin{equation}}
\newcommand{\eeq}{\end{equation}}
\newcommand{\bea}{\begin{eqnarray}}
\newcommand{\eea}{\end{eqnarray}}

\renewcommand{\vec}[1]{\mbox{\boldmath$#1$\unboldmath}}

\def\lsi{\raise0.3ex\hbox{$<$\kern-0.75em\raise-1.1ex\hbox{$\sim$}}}
\def\gsi{\raise0.3ex\hbox{$>$\kern-0.75em\raise-1.1ex\hbox{$\sim$}}}

\begin{document}

\title{On interpretation of fluctuations of conserved charges at high T}

\author{T.~D.~Cohen\inst{1} \and  L.~Ya.~Glozman\inst{2}}
%
\institute{Department of Physics and Maryland Center for Fundamental Physics, University of Maryland,\\ College Park, MD 20742 USA\and
Institute of Physics, University of Graz, A-8010 Graz, Austria}
%
%
\abstract{
Fluctuations of conserved charges calculated on the lattice which can be measured experimentally, are well reproduced by 
a hadron resonanse gas model at temperatures
below $T_{ch} \sim 155$ MeV and radically deviate from the hadron resonance gas predictions
above the chiral restoration crossover. This behaviour is typically interpreted
as an indication of deconfinement in the quark-gluon plasma regime. We present an argument  that
this interpretation may be too simple.  The argument is  based on the scaling of quantities with the number of colors: demonstration of deconfinement and QGP
requires observable that is sensitive to $\sim N_c^2$ gluons  while the conserved charges
are sensitive only to quarks and above  $T_{ch}$ scale as $N_c^1$. The latter scaling  is consistent with the existence of an intermediate regime   characterized by  restored chiral symmetry and by approximate chiral spin symmetry which is a symmetry of confining interaction. In this regime the energy density, pressure and entropy density
scale as $N_c^1$. In the large $N_c$ limit this regime might become a distinct phase separated from the hadron gas
and from QGP by phase transitions. A natural observable that associates with deconfinement and is directly sensitive to deconfined $N_c^2-1$ gluons is the Polyakov
loop; in the $N_c=3$ world it remains very close to 0 at temperatures well above chiral crossover, reaches the value $\sim 0.5$ around
$\sim 3T_{ch}$ and the value  close  to 1 at temperatures $\sim 1$ GeV.
} 
\maketitle

\section{Introduction}	
It was widely believed for a long time that in QCD upon heating
there are two qualitatively different regimes connected by a fast
but smooth crossover: a hadron gas and a deconfined quark-gluon
plasma. The hadron gas, which is a dilute system of hadrons at temperatures below
a critical one, is characterized 
by a spontaneously broken chiral symmetry while the QGP is a system of effectively deconfined
(quasi)quarks and (quasi)gluons with restored chiral symmetry. The pseudo-critical temperature
was believed to be a common temperature of the chiral restoration and of the deconfinement. In 
the real world $N_c=3$ and with small but nonzero quark masses the pseudocritical
temperature of chiral restoration was established to be around $T_{ch} \sim 155$ Mev
\cite{A2,ch1,ch2} while in the chiral limit the critical temperature of the second
order chiral restoration phase transition was determined to be around 130 MeV \cite{Karsch}.
In the large $N_c$ limit it was believed that the common deconfinement and chiral restoration
phase transition is of the first order \cite{mlp}.

Lattice studies on artificial truncation of the near-zero modes of the
Dirac operator at T=0 \cite{D1,D2,D3,D4} have suggested approximate emergent symmetries
associated with the symmertries of the color charge and of the electric part of the QCD Lagrangian \cite{G2,G3}: the $SU(2)_{CS}$ chiral spin symmetry that includes as a subgroup
the $U(1)_A$ symmetry and its flavor extension $SU(2N_F)$.  The latter contains
as a subgroup the full chiral symmetry of QCD $SU(N_F)_L \times SU(N_F)_R \times U(1)_A$.
Neither chiral spin symmetry nor its flavor extension are symmetries of
the Dirac Lagrangian but are symmetries of only the electric  part of the QCD Lagrangian; they
are explicitly violated by the magnetic interactions and by the quark kinetic terms. For a review
on symmetries and their implications for hot QCD see Ref. \cite{G1}. 

Above the chiral symmetry restoration crossover one
might expect emergence of the approximate $SU(2)_{CS}$ and $SU(4)$ symmetries \cite{G4}, which
would suggest that the system is still in the confining regime. 
Lattice studies appear to be consistent with this \cite{R1,R2,R3,Chiu}.
Namely at temperatures roughly  $T_{ch} < T < \sim 3T_{ch}$ one observes correlators consistent with approximate 
chiral spin symmetry and its flavor extension; this behaviour smoothly disappears at higher temperatures.
This suggests that in QCD upon heating above the hadron gas regime but below the quark-gluon
plasma regime there exists an intermediate confining regime with restored chiral symmetry and approximate chiral spin symmetry.

There are arguments about the dynamics about how this intermediate regime behaves and how its effective degrees of freedom arise\cite{G1} but for the purpose of this paper the validity of these arguments is not relevant.

The  center symmetry of the gauge 
action is explicitly broken by quark loops.  However, it becomes exact in the large $N_c$ limit\footnote{There is a subtlety in that the large $N_c$ limit should to be taken at the end of the analysis in order to maintain hierarchies associated with factors $N_c$ to various powers. If the large $N_c$ limit is taken at the outset, as in Ref. \cite{Kovtun:2007py}, there is a possibility that explicit center-symmetry-breaking effects associated with quarks which could conceivably enter at  order $N_c$ in the action ({\it i.e.} relative order $1/N_c$) would be missed.  The scenario discussed in this paper assumes that it happens.}: QCD  in the combined large $N_c$ and chiral limit has two distinct symmetries: chiral
symmetry and center symmetry.  This allows one to  define unambiguously possible phases with confinement or
deconfinement and with spontaneously broken or restored chiral symmetry. The 
order parameters for the two  (Polyakov loop, quark condensate) are independent. It was suggested in Ref. \cite{CG}
that three regimes of QCD connected by smooth crossovers in the real world $N_c=3$ and with small
but nonzero quark mass might become distinct phases separated by phase transitions once the
number of colors increases to infinity with the quarks kept massless. Standard large $N_c$ scaling analysis implies that the energy density, pressure and entropy density in the hadron gas phase scale as $N_c^0$, in the intermediate phase they scale as $N_c^1$
and in the QGP phase as $N_c^2$.
 The intermediate phase, which is
chirally symmetric and confined, has a temperature range  $\sim N_c^0$ and should be at least approximately chiral spin symmetric. It should contain
a gas of noninteracting glueballs  (as is seen in the low temperature hadron gas phase). 

The confined
and chirally symmetric phase at $T=0$ and large density  was discussed in ref. \cite{mlp}. In that case the chiral symmetry restoration can be attributed to a large quark Fermi sphere.  The  mechanism of  chiral symmetry restoration at large $T$ and vanishing chemical potential is  clearly different; ref. ~ \cite{gnw} argues that it is due to Pauli blocking of the quark
levels, necessary for the formation of the quark condensate, by the thermal
excitations of quarks and antiquarks .

 There are models - both quark-based and meson-based, that effectively imitate
 confinement in thermodynamics, via a coupling of the light quark
 sector to the effective Polyakov loop potential \cite{Pis,Fuk,Rat,lrs}.
 Within these models, at vanishing chemical potential the chiral restoration phase transition at large $N_c$ typically coincides or very
 close
 with the "deconfinement" phase transition.  This behaviour is notably different from  what is proposed here
based on large $N_c$ considerations.

The above overview of  developments of the field   introduces  the main subject of this paper, that will be discussed in the following sections: interpretation of the fluctuations of conserved
charges related to quark bilinears. 
 We stress that the  purpose of the paper is not to calculate
something new but rather to give a correct interpretation to
already existing results that are rather influential for the understanding
of the physics at high temperatures.

\section{Conserved charges and their fluctuations in hadron gas and at higher temperatures}

The hadron resonance gas model of the QCD matter at low temperatures
assumes a dilute system  of point-like structureless hadrons
that do not interact. Consequently the number density of meson or baryon species $k$  with spin $S_k$,
isospin $I_k$ and  some  strangeness  at a temperature $T$ is given by the Bose-Einstein
distribution (minus sign in the denominator of the equation below) for mesons and
Fermi-Dirac (plus sign) for baryons

\begin{equation}
  n_k(T)  = (2S_k+1)(2I_k+1) \int \frac{{\rm d}^3 p}{(2 \pi)^3} \, \frac{1}{e^{\sqrt{p^2 + m_k^2}/T} \pm 1}  \; .
\end{equation}
The  fluctuations of   the numbers of the various quark flavors are then 
obtained from the standard results   for the Bose- and Fermi-gases at the
given $T$ assuming hadron masses from PDG. It is well known that for $T < T_{ch}$ the hadron resonance gas model  reproduces 
fluctuations of conserved charges calculated on the lattice, see e.g. \cite{Bel} and references therein.  At larger temperatures the lattice results
for these observables  radically deviate from the HRG model predictions. This is often taken
as evidence of deconfinement and of transition to QGP.  However as it will
become evident below such interpretation is erroneous.

The success of the HRG model below $T_{ch}$ indicates that at these temperatures
the hadron structure is not yet resolved and its internal degrees of freedom
are frozen. This is the reason for the $N_c^0$ scaling of the thermodynamical
observables in the HRG regime.
Once the density of hadrons increases so that they start to overlap the internal hadron structure gets relevant and a proper description of conserved charges should rely on quark degrees of freedom.

Here we focus on charges associated with the net number of up, down and strange quarks\footnote{It is sensible to use these rather than the more traditional baryon number and electric charge as well as strangeness, since our goal  is to make connection with the behavior at large $N_c$ and the extrapolation to large $N_c$ is straightforward and unambiguous with these and not with elecric charge or baryon number. There are two distinct ``natural'' ways to define the electric charge of quarks:  $Q_u=\frac{2}{3}$ and $Q_d=Q_s=-\frac{1}{3}$ as at $N_c=3$ or alternatively  \cite{gl,Bar} $Q_u=\frac 1{2N_c}+1/2$ and $Q_d=Q_s=\frac 1{2N_c}-1/2$.   Only the second definition gives the proton ($B=1$, $I=1/2$, $I_3=1/2$ and $S=0$)  electric charge of unity, while the first has the proton charge  of order $N_c$.  Similarly, the natural definition of baryon number has the baryon number of a quark equal to $1/N_c$  rather than $\frac{1}{3}$ which gives rise to suppression factors $1/N_c$ that can obscure the physics. }
:
\begin{equation}    
N_q \equiv \int d^3 x ~n_q(x) \; \;\;  {\rm with}
\; \; \;n_q(x) = \bar q(x) \gamma^0 q(x), \;\; \; q=u,d,s
\label{def}
\end{equation}

Each quark can be in one of the  $N_c$ color states and contraction with respect
to the color of quarks is assumed. This means that the  conserved
flavor charges  $N_q$, scale as $N_c^1$.  Then the fluctuations of
the conserved quark charges also scale as $N_c^1$.

 To see this note that the expectation value of the conserved quark number of a given flavor in  volume $V$ at temperature $T$ can be
obtained from the grand canonical partition function as

\begin{equation}
<N_i> = \frac{T \partial \left[log Z(T,V,\mu_u,\mu_d,...)\right]}{\partial \mu_i}.
\end{equation}
The fluctuations of conserved charges  can be obtained as a derivative
of these charges

\begin{equation}
\frac {\partial <N_i>}{\partial  \mu_j} = \frac{T \partial^2 \left[log Z(T,V,\mu_u,\mu_d,...)\right]}{\partial \mu_j \partial \mu_i}.
\end{equation}
The fluctuations and correlations of conserved charges can be expressed in terms
of different cumulants

\begin{equation}
\chi_{i,j,k}^{u,d,s} = \frac{T \partial^{i+j+k} \left( P/T^4\right)  }{(\partial\mu_u)^i (\partial \mu_d)^j  (\partial \mu_s)^k }.
\end{equation}

In Figs. 1 and 2 we show typical results for fluctuations of quark numbers 
of $u,d,s$ quarks taken from Ref. \cite{Bel} and their comparison with the
HRG model. We see
that the fluctuations of the $u,d,s$ quark numbers deviate from
the HRG just at the chiral restoration temperature 155 MeV. 

\begin{figure}[h]
\centering
 \includegraphics[width=0.5 \textwidth]{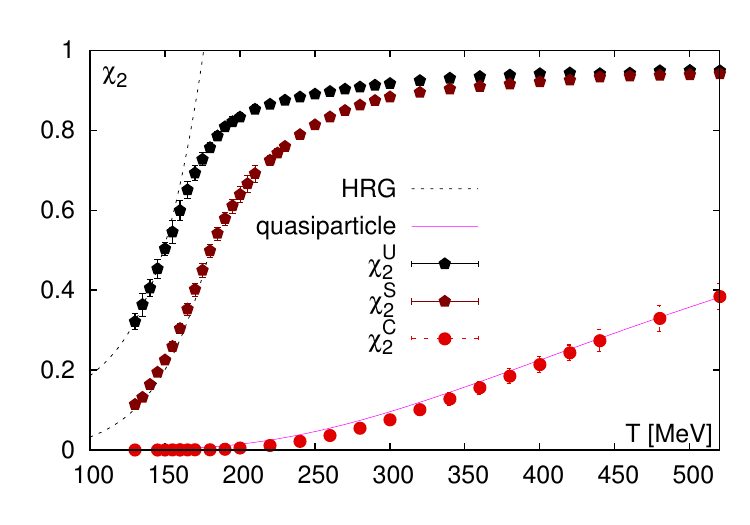}
    \caption{Fluctuations of conserved net $u$ ($\chi_2^U= \chi_{2,0,0}^{u,d,s}$) and strange ($\chi_2^S= \chi_{0,0,2}^{u,d,s}$) quark numbers in 2+1 QCD at physical quark masses. $\chi_2^C$
    is irrelevant to our discussion and can be ignored. Source: Ref. \cite{Bel}.}
    \label{fig1}
\end{figure}

\begin{figure}[h]
\centering
 \includegraphics[width=0.5 \textwidth]{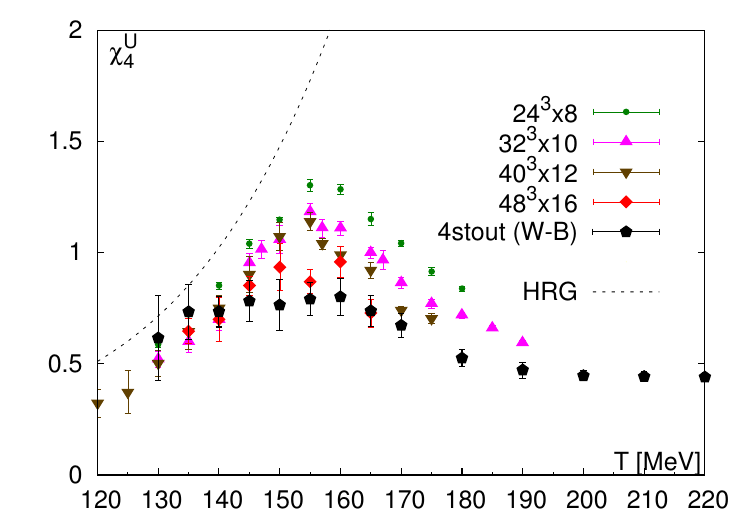}
    \caption{Cumulant $\chi_4^U= \chi_{4,0,0}^{u,d,s}$  in 2+1 QCD at physical quark masses. Source: Ref. \cite{Bel}.  }
    \label{fig2}
\end{figure}

The key point of this paper is that since the quark numbers scale as $N_c^1$ the above behaviour of
fluctuations of conserved charges is consistent with the crossover from the hadron gas to the  intermediate regime, as described in the introduction,
where all main thermodynamical quantities scale as $N_c^1$. Consequently
above $T_{ch}$ the fluctuations of conserved charges in the large
$N_c$ limit
 would be expected to differ from the fluctuations in the HRG (of order $N_c^0$) by $N_c$, indicating a phase transition.  In principle 
the scaling of the fluctuations should be seen on the lattice  by providing
calculations at $N_c > 3$. However,the $N_c^1$ scaling is a robust consequence of the
 definition of conserved charges (\ref{def}) and hence there is no
 reason to doubt this scaling.

To demonstrate a transition to the
QGP regime one needs an observable that is sensitive to presence of $\sim N_c^2$ deconfined  gluons.

\section{Polyakov loop and its evolution with temperature}

The Polyakov loop is the trace of a Wilson line along a straight path
in the  compactified time direction 

\begin{equation}
 P_{N_c} = \frac{1}{N_c} Tr \left[T \exp{\left(i \int_0^\beta d \tau A_0(\vec{x},\tau)  \right)}  \right].
 \end{equation}

\begin{figure}[h]
\centering
 \includegraphics[width=0.5 \textwidth]{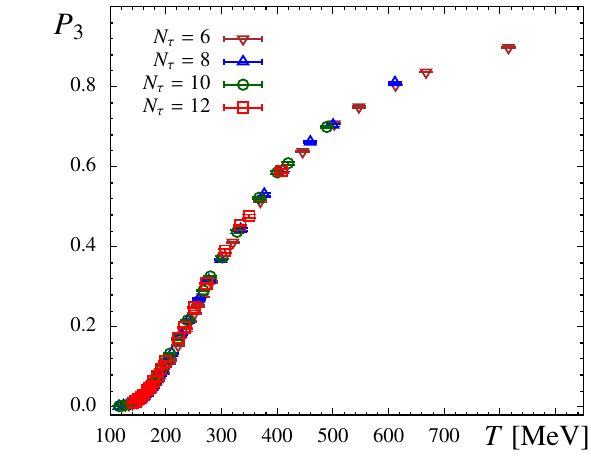}
    \caption{Temperature evolution of the properly renormalized Polyakov loop in 2+1 QCD at physical quark masses. Source: Ref. \cite{PS}}
    \label{fig3}
\end{figure}

 The expectation value of the Polyakov loop in the pure glue theory (or in quenched QCD and in QCD
 with infinitely heavy quarks) is the order parameter for center symmetry and for confinement. In the center
 symmetric confining phase at low temperatures the expectation value of the Polyakov loop identically vanishes while above the first order deconfinement phase transition  the center symmetry gets spontaneously broken and the expectation value of the Polyakov loop takes a finite value.

It is important to clarify the $N_c$ properties of the Polyakov loop.
The factor $\frac{1}{N_c}$ in front of the trace   normalises the trace of the
$N_c \times N_c$  matrics. The absolute value of the normalised Polyakov loop varies in the
interval [0,1] and when it is nonvanishing scales as $N_c^0$.

 The Polyakov loop in the
deconfined phase, where it is not zero, is explicitly sensitive to $ N_c^2-1$
gluons, which is the dimension of the adjoint representation, i.e. the number
of independent gluons.  The sensitivity should not be confused with the scaling.

 It is known that the temperature of the first order  deconfinement phase transition in pure Yang-Mills and in quenched QCD at $N_c=3$ is at $T_d \sim  270 - 300 $ MeV. At this temperature
 the Polyakov loop jumps from the zero value to the value around 0.5 - 0.6 and above the phase transition smoothly increases towards 1. It is also established that the deconfinement temperature in
 pure Yang-Mills theory is practically $N_c$-independent \cite{ln2}. However, confining
 properties of QCD with light quarks
 and of pure Yang-Mills theory are identical in the large $N_c$ limit. Then one expects  a similar
 deconfinement temperature of  QCD with light quarks at large $N_c$.

 In the real world $N_c=3$ in QCD with light quarks the center symmetry of the action is explicitly broken and 
the deconfinement first order phase transition 
 is replaced by a very smooth crossover.  The renormalized
Polyakov loop informs  us about the deconfinement crossover region.

The temperature evolution of the properly renormalized Polyakov loop, taken from Ref. \cite{PS}, is shown in Fig. 3. We observe that  above the chiral restoration temperature around
$T_{ch} \sim 155$ MeV the Polyakov loop is very small which suggests that here QCD is
 in the confining regime. At the same time at these temperatures fluctuations of
 conserved charges demonstrate that the hadron gas picture does not work. Both these
 facts  are consistent with the existence of the intermediate regime discussed in the introduction. The
 Polyakov loop reaches the value around 0.5 at a temperature roughly $~3 T_{ch}$, in agreement
 with the temperature of smooth disappearance of chiral spin symmetry.
 

 \section{Conclusions}
 In this paper we have considered fluctuations of conserved charges which are
 typically taken as evidence for transition from hadron gas to a QGP at the
 chiral restoration temperature. As the  $N_c$ scaling analysis makes manifest, the transition to the QGP requires, however, an observable
 that is sensitive to the presence of deconfined gluons. 
 A natural observable is the Polyakov loop.
 The conserved charges associated with quark flavor scale as $N_c$ and are not explicitly dependent  on the presence of deconfined
 gluons. The fluctuations and correlations of conserved charges measured on the lattice
 indicate a transition from the hadron gas to a regime with the scaling $N_c$. At the
 same time the Polyakov loop just above the pseudocritical temperature of chiral
 restoration remains very small suggesting a confining regime. This data supports
 the previously found evidence that above the HRG regime QCD matter is in not the
 deconfined  QGP regime, but in the confined regime with restored chiral symmetry
 and approximate chiral spin symmetry. A very smooth transition to the QGP regime from the
 intermediate regime takes place at essentially larger temperatures.

\section*{Acknowledgements}
  The work of TDC was supported in part by the the U.S. Department
of Energy, Office of Nuclear Physics under Award Number
DE-FG02-93ER40762.

\end{document}